# Application of an attention-based CNN-BiLSTM framework for *in vivo* two-photon calcium imaging of neuronal ensembles: decoding complex bilateral forelimb movements from unilateral M1

Ghazal Mirzaee[1], Jonathan Chang[2], Shahrzad Latifi[3*]

[1]Lane Department of Computer Science and Electrical Engineering, West Virginia University, Morgantown, WV, USA

[2]University of Miami Leonard M. Miller School of Medicine, Miami, FL, USA

[3]Department of Neuroscience, Rockefeller Neuroscience Institute, West Virginia University, Morgantown, WV, USA

**\*Corresponding author**

S.Latifi  sl00092@hsc.wvu.edu

## Abstract

Decoding behavior, such as movement, from multiscale brain networks remains a central objective in neuroscience. Over the past decades, artificial intelligence and machine learning have played an increasingly significant role in elucidating the neural mechanisms underlying motor function. The advancement of brain-monitoring technologies, capable of capturing complex neuronal signals with high spatial and temporal resolution, necessitates the development and application of more sophisticated machine learning models for behavioral decoding. In this study, we employ a hybrid deep learning framework, an attention-based CNN-BiLSTM model, to decode skilled and complex forelimb movements using signals obtained from *in vivo* two-photon calcium imaging. Our findings demonstrate that the intricate movements of both ipsilateral and contralateral forelimbs can be accurately decoded from unilateral M1 neuronal ensembles. These results highlight the efficacy of advanced hybrid deep learning models in capturing the spatiotemporal dependencies of neuronal networks activity linked to complex movement execution.

# Application of an attention-based CNN-BiLSTM framework for *in vivo* two-photon calcium imaging of neuronal ensembles: decoding complex bilateral forelimb movements from unilateral M1

## INTRODUCTION

Brain is a complex system of networks across multiple spatiotemporal scales. These highly optimized interconnected networks shape the brain as the engine of cognition and movement. Among multiscale brain networks, ranging from networks of genes at nanoscale to brain regions at macroscale, mesoscale networks play pivotal roles in signaling and information processing. In the brain, a mesoscale neural network is defined as the level between individual neural cells to neural ensembles and circuits[1,2]. In recent years, advances in brain-tracking technologies have significantly enhanced the ability in real-time monitoring of detailed mesoscale brain networks activity. Among these, *in vivo* two-photon calcium or voltage imaging provide the possibility of extracting signals from individual neurons of a broad neuronal population in the living brain and during specific behavior[3]. This technology has been widely used in recent studies of neuronal circuits underlying behaviors such as motor control and movement.

The primary motor cortex (M1) plays a crucial role in the acquisition and control of new and complex motor skills. In fact, it has been shown that the activity of neuronal networks in M1 is correlated with a range of movement-related variables such as direction, speed of movement, or accurate placement of limbs during locomotion[4]. Encoded information from M1, therefore, can be used as sources for neural prostheses in paralyzed cases to regain movement in lost arm/limb[5]. M1 impacts motor output by orchestrating sensory-guided movement coordination and navigational computations, facilitating required adjustments in gait[6-8]. Consequently, M1 lesion, in acquired brain disorders such as ischemic stroke or traumatic brain injury (TBI), can result in locomotor impairments such as gait deficits[9]. M1 exhibits lateralized outputs, predominantly controlling movements on the contralateral side (e.g. arm)[10]. However, studies in macaques have demonstrated that M1 neurons can be active during movement of either arm, particularly in bimanual tasks or during interactions that demand bilateral coordination[11]. Despite these findings, it remains uncertain to what extent computations of unilateral M1 neurons are involved in bilateral movement, and if specific neuronal ensembles contribute more to this process. This raises the question regarding the ability of unilateral neuronal networks in M1 to encode information sufficient to induce complex movements on both sides of the body, a particularly critical consideration for individuals with brain injury and paralysis, such as those resulting from war-related trauma.

In this study, we investigated whether complex motor movements in both forelimbs of a mouse can be decoded from the activity of excitatory neuronal ensembles in unilateral M1. We developed a multimodal approach that integrates neuronal network activity, ipsilateral/contralateral forelimb motor movements, and deep learning models. Advanced *in vivo* 2P calcium imaging was used to extract neuronal networks data from the activity of excitatory neuronal ensembles in unilateral M1. A motorized grid-walking setup was built to induce coordinated complex forelimb motor movement of gait in ipsi- and contralateral limbs as described previously[12]. To decode behavior from neuronal network activity, we first recruited recurrent neural networks (RNN); A Long short-term memory

(LSTM) with fully connected layers, was implanted to capture the temporal dependencies inherent in the time-series neuronal signals[13, 14]. Since the temporal modeling alone did not account for the spatial dependencies among neurons and their coordinated influence on complex forelimb motor movement, we developed an advanced hybrid architecture that combined Convolutional Neural Networks (CNNs), Bidirectional LSTM (BiLSTM) networks, and attention mechanisms[15]. This novel architecture proved to be robust and accurate, effectively learning the intricate mapping between neuronal activity patterns and their corresponding motor outputs of ipsi- and contralateral forelimbs during complex gait movement. Our study demonstrated that: 1) excitatory neuronal ensembles (at the single neuron level) in unilateral M1 can be leveraged to predict complex movement in bilateral forelimb using deep learning models, 2) application of hybrid deep learning model enhances the decoding accuracy of gait movement based on the neuronal signal extracted from 2P calcium imaging highlighting the importance of spatial features of neuronal networks in decoding. These data suggest that hybrid and more advanced deep learning models hold significant promises for improving brain-inspired neural decoding technologies by enhancing our understanding of neuronal network computation and motor control. Furthermore, the ability to decode movement in both forelimbs from the networks of a single hemisphere at the neuronal population level, highlights the potential for developing more sophisticated and integrated models in neuroprosthetics and broader applications in brain computer interface (BCI).

## RESULTS

### Integrating skilled forelimb motor task with *in vivo* two-photon calcium imaging and deep learning

To assess the activity of excitatory neuronal networks in M1, a CaMKII-tTA/tetO-GCaMP6s transgenic line was employed[3]. This line provides persistent and uniform expression of genetically encoded calcium indicators (here GCaMP6s) in excitatory neurons. A 5 mm cranial window was stereotaxically placed on the head of adult male mouse to provide 2P imaging from layer II/III of unilateral forelimb M1. A head-fixed motorized grid-walking paradigm was developed to simultaneously track ipsi- and contralateral forelimb movements under the 2P microscopy[12] (Figure 1). The grid-walking task evaluates forelimb placement on a complex grid during exploratory gait and is commonly used to detect skilled-motor control, such as coordinated gait, as well as motor deficits like foot-fault in brain disorders (e.g., stroke or TBI)[16]. We defined a complex motor movement in ipsilateral and/or contralateral forelimb as a complete footstep, characterized by continuous movement beginning with the release of the paw from the initial grid and ending with reaching and grasping the subsequent grid by the corresponding paw (Figure 1B). The complete footstep on the motorized grid-walking wheel with constant velocity displayed both rhythmic and discrete movement. After training the animal on the grid-walking wheel for 4 days, recording sessions were conducted at 15 Hz frequency, capturing a total of 3000 frames per session; motor movements of the ipsi- and contralateral forelimbs were evaluated in behavioral videos, with 2P recording frames

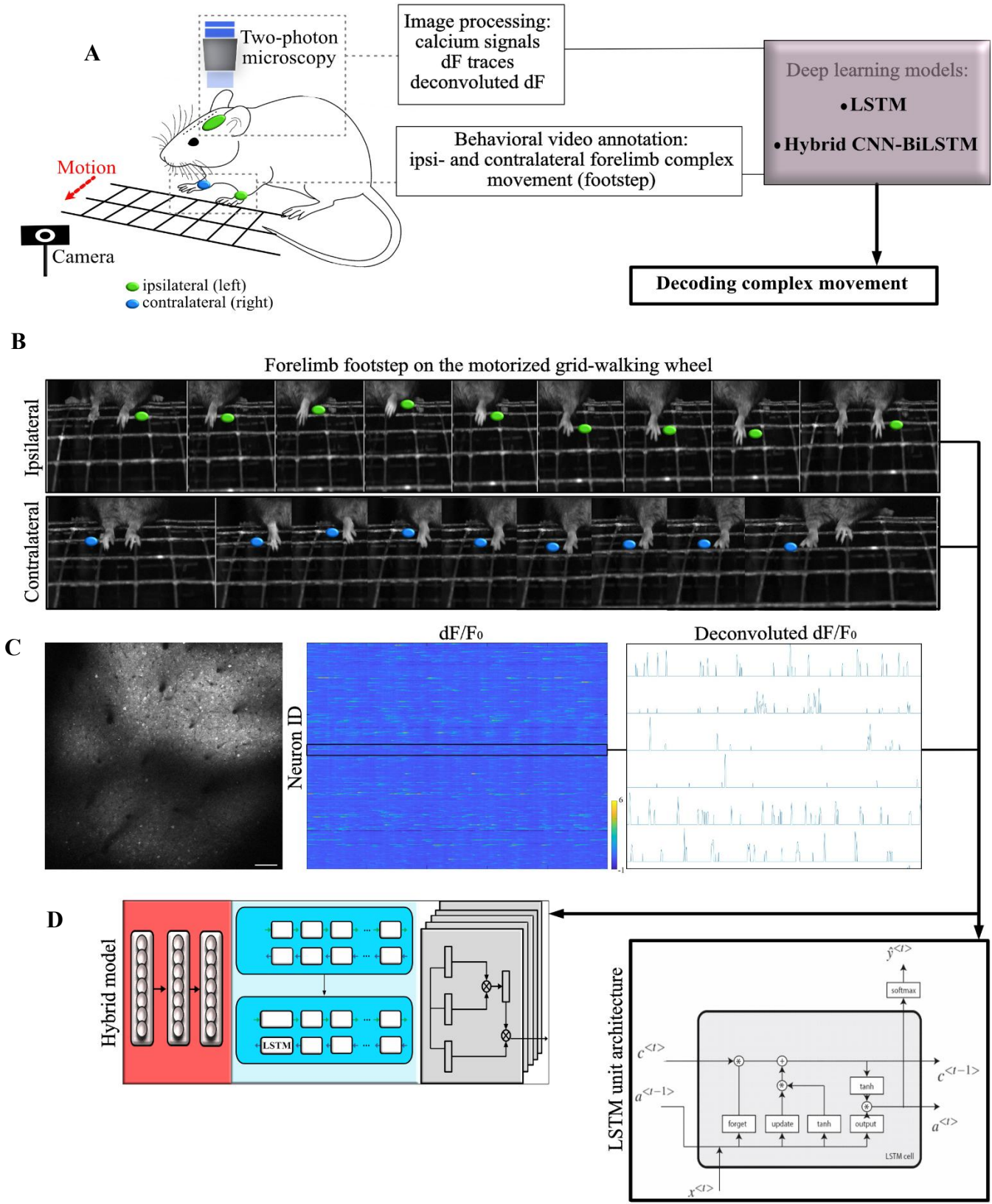

**Figure 1.** Representative image of experimental setups and analysis approaches. A) 2P microscopy is used to track M1 neuronal activity while an animal performing complex motor behavior on the motorized grid-walking wheel. B) An example of ipsi- (upper panel) and contralateral (lower panel) forelimbs complex movements on the behavioral setup. A forelimb complex movement is defined as a forelimb footstep on the motorized grid-

walking wheel characterized by continuous movement beginning with the release of the forelimb paw from the initial grid and ending with reaching and grasping the subsequent grid by the corresponding paw. Ipsi- and contralateral forelimb footsteps are evaluated from behavioral videos annotation and the corresponding 2P recording frames are labeled accordingly. C, left panel: an example of one frame from recorded field of view scale bar 100 μm). A series of image processing algorithms are applied to extract dF/F0 traces (C, middle panel) and then to compute deconvoluted dF/F0 (C, right panel) signals from recorded 2P videos (each video: 3000 frames, 15 Hz rate). D) Two deep learning models (LSTM and CNN-BiLSTM) are employed to decode bilateral forelimbs complex movements from 2P imaging data.

annotated to correspond to each footstep. To minimize motion artifacts caused by animal movement on the motorized wheel during recording, the motion correction algorithm (NoRMCorre) was applied to the recorded videos[17]. The corrected videos underwent image processing to first identify regions of interest (ROIs) representing neurons, and extract calcium transients for each ROI. Further analyses were then applied to compute $dF/F_0$ values $[(F(t) – F_0)/F_0]$ for each ROI[12]. Post-processing algorithms (deconvolution) have been used to mitigate the slow decay issue associated with GCaMP6s[12, 18] (Figure 1C). The achieved data was normalized to ensure consistency in feature scaling, splitting into training, validation, and test sets using sampling to preserve class distribution. To capture complex temporal dependencies in neuronal activity sequences, we first implemented a LSTM-RNN; LSTMs are type of RNN, containing a special cell state and gating mechanism that allows them to selectively remember or forget information from previous time-steps[14]. The deconvoluted neuronal signals were used to feed the LSTM as sequence input, enabling the model to learn the temporal dynamics and relationships between frames. (Figure 1D). A multi-class classification task was formulated as follows: 0 for no footstep, 1 for contralateral (right) footstep, and 2 for ipsilateral (left) footstep. A sequence of labels corresponding to the sequence of frames was achieved during the evaluation of model predictions. Consecutive frames with the same label (e.g., 1 for contralateral footstep) form a continuous segment representing the interval in which the footstep with the ipsilateral or contralateral forelimb was performed. Further post-processing was applied to the predicted labels to extract the start and end frames of the footstep intervals. For example, the first and last consecutive frames with label 1 were identified to determine the start and end of the footstep intervals with the right foot. By leveraging the sequential nature of LSTMs and the multi-class classification approach, we obtained predictions for individual frames and, at the same time, captured the intervals or segments of frames corresponding to footsteps. This approach provides more detailed information compared to simply predicting individual footstep frames. However, the model's performance relies not only on the quality and representativeness of the training data but also the ability of the extracted neuronal activity to accurately capture the complex dynamics of footsteps. Therefore, as mentioned before, we employed deconvoluted neuronal signals to persevere the rapid neuronal events.

**First model: a fully connected LSTM to decode complex movements of ipsi- and contralateral forelimbs from unilateral M1**

To capture complex temporal dependencies in neuronal activity sequences, we first implemented a LSTM-RNN; LSTMs are type of RNN, containing a special cell state and gating mechanism that allows them to selectively remember or forget information from previous time-steps[14]. To preserve rapid neuronal events, the deconvoluted neuronal signals were used to feed the LSTM as sequence input, enabling the model to learn the temporal dynamics and relationships between frames (Figure 1D).

We implemented three classification tasks: binary classification for ipsilateral (left) forelimb footstep, binary classification for contralateral (right) forelimb footstep, and a multi-class classification task formulated as follows: 0 for no footstep, 1 for contralateral footstep, and 2 for ipsilateral footstep. A sequence of labels corresponding to the sequence of frames was achieved during the evaluation of model predictions. Consecutive frames with the same label (e.g., 1 for contralateral footstep) form a continuous segment representing the interval in which the footstep with the ipsilateral or contralateral forelimb was performed. Further post-processing was applied to the predicted labels to extract the start and end frames of the footstep intervals. For example, the first and last consecutive frames with label 1 were identified to determine the start and end of the footstep intervals with the contralateral forelimb. By leveraging the sequential nature of LSTMs and the multi-class classification approach, we obtained predictions for individual frames and, at the same time, captured the intervals or segments of frames corresponding to footsteps. This approach provided more detailed information compared to simply predicting individual footstep frames.

| Hyperparameter | Value |
| --- | --- |
| Learning Rate | 0.0001 |
| Batch Size | 64 |
| Epochs | 30 |
| Optimizer | Adam |
| Loss Function | Cross Entropy |
| Regularization | L2 (1e-5) |
| Dropout Rate | 0.5 |
| LSTM Configurations | Input→LSTM (hidden_size=64, num_layers=2)→ BatchNorm |
| Fully Connected Layers | 64→32 |
| Activation Functions | ReLU |
| Learning Rate Scheduler | ReduceLROnPlateau (factor=0.5, patience=5) |
| Gradient Clipping | Max norm = 1.0 |
| Weight Initialization | Xavier/Orthogonal (layer dependent) |

**Table 1.** Hyperparameters used for fully connected LSTM model.

Our proposed LSTM model included standardized neuronal signal (deconvoluted) inputs with the behavioral task labels as output targets. The neural signal input was passed through an LSTM layer with 2 layers and configurable hidden size (64 units used in experiments), followed by batch normalization and fully connected layers with 64 and 32 units, respectively, with dropout between layers to prevent overfitting (See Method and Table 1). The output layer was a dense layer with softmax activation providing the predicted footstep status (2 units for binary tasks, 3 units for multiclass).

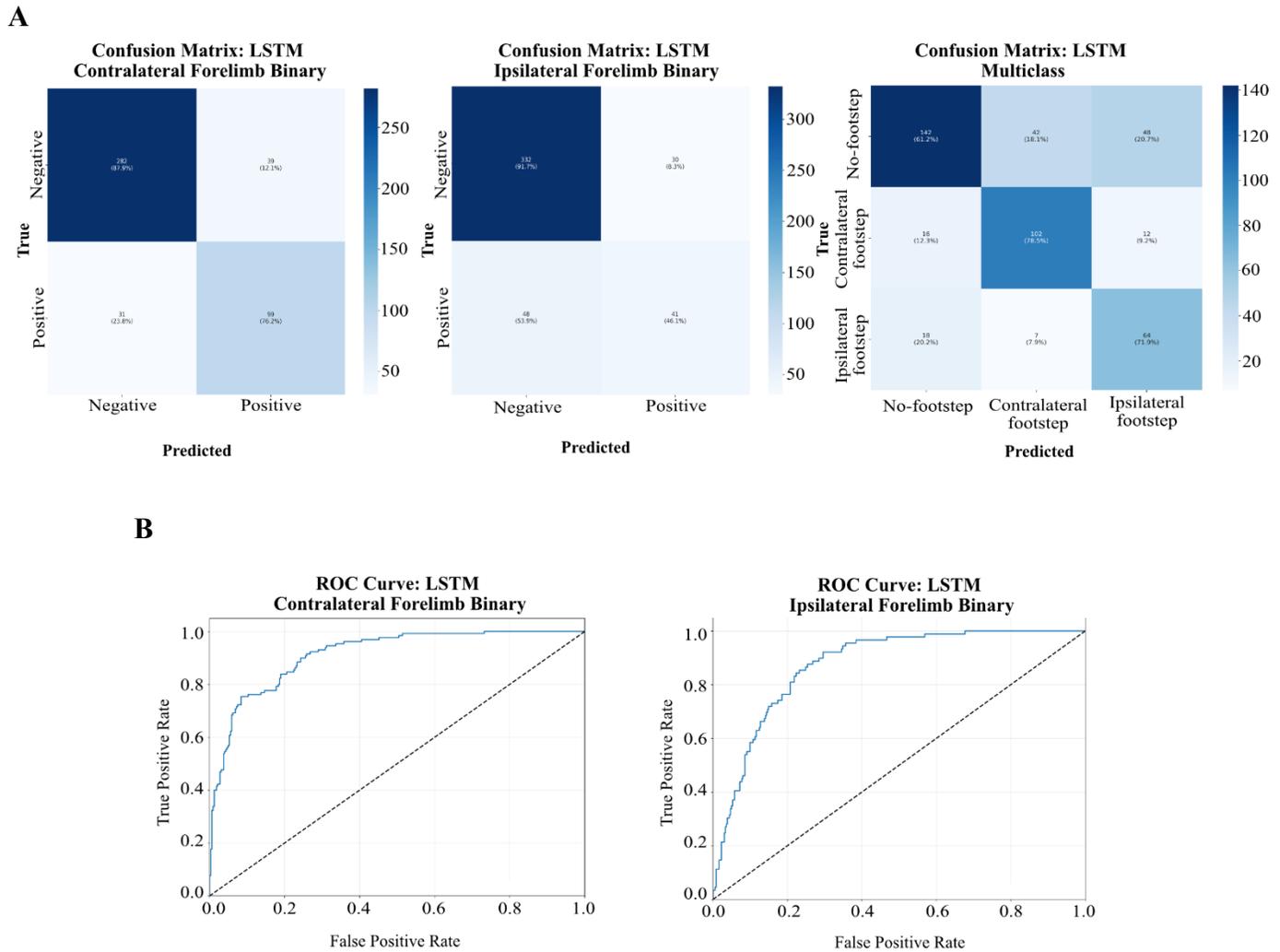

**Figure 2. LSTM model performance in classifying forelimbs complex movements. A)** Confusion matrices for LSTM model in decoding forelimbs footsteps on the grid-walking wheel. The multiclass confusion matrix (right) highlights good accuracy in classifying no-footstep states (142 out of 232 correctly identified) and distinguishing ipsilateral (332 correct) and contralateral footsteps (282 correct). The binary classifiers (center and left) demonstrate improved discrimination for single-forelimb detection, with contralateral and ipsilateral footstep classifiers achieving 282/99 and 332/41 true negatives/positives, respectively. The binary classifiers show notably higher performance for ipsilateral foot detection compared to contralateral movements, suggesting asymmetric neuronal representation in the recorded population. These results validate the model's ability to extract meaningful bilateral forelimb motor patterns from neuronal activity, while revealing differential encoding strength between forelimbs. **B)** ROC Curves for LSTM Binary Classification of contralateral (left panel) and ipsilateral (right panel) footsteps. Both ROC curves indicate strong discriminative ability, with high true positive rates (TPR) at low false positive rates (FPR).

The fully connected layer was added to enable the model to learn complex non-linear relationships between the extracted features (which are the temporal patterns in neural activity captured by the LSTM layers) and the target variable (the footstep classification labels). The model was trained using the Adam optimizer with learning rate scheduling and early stopping to prevent overfitting. Class weighting was incorporated in the loss function to address imbalance in the data. For decoding performance several metrics were evaluated, including accuracy, precision, recall, and F1 score. The performance metrics in Table 3 demonstrate the model's ability to predict the complex movements of bilateral forelimbs. The highest accuracy of 84.4% was achieved for contralateral footstep prediction, with an F1 score of 73.8% (Table 3). In comparison, the metrics for ipsilateral prediction were 82.7% and 51.2%, respectively (Table 3). For multiclass prediction, defined as simultaneous ipsilateral, contralateral, and no footstep classification, the performance metrics were 68.2% for accuracy and 68.5% for F1 score. Summary of the models' performance from confusion matrix (in terms of true positive, true negative, false positive, and false negative predictions) is presented in Figure 2. For contralateral forelimb footstep detection, the model correctly identified 99 movement events and 282 non-movement periods, with 39 false positives and 31 false negatives, demonstrating strong classification performance particularly in identifying true movement events. Similarly, for ipsilateral forelimb movement prediction, the model successfully detected 41 movement events and 332 non-movement periods, with 30 false positives and 48 false negatives, showing performance in distinguishing movement states. In the multiclass prediction task, where the model simultaneously classified no movement, ipsi- and contralateral forelimb footsteps, the confusion matrix revealed good performance characteristics. The model achieved the highest accuracy in identifying no-movement states (142 correct classifications), while also showing strong performance in distinguishing between contralateral forelimb (102 correct classifications) and ipsilateral forelimb (64 correct classifications) footsteps. The confusion patterns indicated interactions between movement types, suggesting the model is more conservative in movement detection in the multiclass setting. The ROC curves provided additional insight into the model's discrimination capabilities (Figure 2). For contralateral footstep detection, the curve showed strong classification performance with a clear deviation from the random classifier line (diagonal), particularly in the early phase of the curve where the true positive rate rose sharply with minimal increase in false positives. The ipsilateral movement detection exhibited similarly strong performance, with the ROC curve demonstrating excellent discrimination ability and a rapid rise in true positive rate while maintaining a low false positive rate. Moreover, our results did not reveal a significant difference in decoding performance between the ipsilateral and contralateral forelimbs across all datasets. The best overall performance was observed on dataset 2B, with decoding performance varying across different datasets (Table 3). These variations may suggest the presence of dataset-specific characteristics. Potential contributing factors could include differences in signal quality and/or noise levels, animal performance during the recording sessions, and the dynamics of calcium transients in the field of recording, reflecting neuronal activity. In our experimental design and analysis, we made efforts to minimize these factors. Overall, these data suggest the feasibility of using unilateral M1 neuronal networks to decode complex movement of footsteps on the grid-walking wheel, supporting the potential role of unilateral M1 neuronal activity in contributing to the movement of both forelimbs.

**Optimizing performance metrics of decoding: application of attention-based CNN-BiLSTM model in *in vivo* 2P calcium imaging of M1 neuronal ensembles**

After investigating whether we can decode bilateral forelimb complex movements from unilateral M1 neuronal network activity, our focus was to develop a model with optimized performance accuracy. This was particularly crucial for improving multiclass prediction accuracy in simultaneously distinguishing between contralateral, ipsilateral, and no footsteps. Such improvements are essential for enabling more precise control in applications that require real-time, dynamic responses, such as BMI. The LSTM with fully connected layers employed here was limited to capturing the temporal dependencies in neuronal activities and thereby neglecting the complex spatial relationships among active neurons and their potential coordinated contributions to movement[19]. Neuronal signals exhibit both spatial correlations across different neurons and temporal dependencies across time points[20]. In fact, traditional single-architecture approaches often struggle to capture these multi-dimensional relationships effectively. CNNs excel at detecting local spatial patterns but may miss long-range temporal dependencies. To overcome these issues, we combined these architectures as a hybrid supervised model (CNN-BiLSTM) with attention mechanisms[21]. This architecture aimed to blend the strengths of both data modalities (behavioral and neuronal) and computational networks (BiLSTM for temporal, convolutional for spatial features), enabling simultaneous prediction of our defined complex forelimb movement in animal[22]. The model allowed it to utilize the temporal dependencies captured by the BiLSTM and the complex spatial relationships and correlations among the recorded neuronal networks. The unique design of the hybrid model and its regularization techniques are outlined below, providing a detailed description of whether these components could contribute to enhancing model performance.

Our input data consisted of temporal sequences of neuronal activity recorded from unilateral M1 neuronal networks[22]. Each temporal sequence was represented as a matrix X ∈ R^(T×N), where T = 32 represented the sequence length and N was the number of neurons:

$$X = x_1, x_2, ..., x_T \in \mathbb{R}^{T \times N} \tag{1}$$

Each time-point xt captured a snapshot of neuronal activity across all neurons:

$$x_t = [n_1^t, n_2^t, ..., n_N^t] \tag{2}$$

where nti represented the activity of neuron i at time t[19]. To address the high variability in neuronal firing rates and ensure stable training, our model implemented a hierarchical normalization strategy directly within its architecture[23]. Rather than applying static preprocessing normalization, we used dynamic batch normalization throughout the network. This began in the CNN layers:

$$\hat{x}t, n = \frac{xt, n - \mu_B}{\sqrt{\sigma_B^2 + \epsilon}} \tag{3}$$

where $\mu_B$ and $\sigma_B^2$ were the batch mean and variance computed across the mini-batch, and $\epsilon$ was a small constant for numerical stability[23]. We applied this normalization after each convolutional layer, with the dimensions progressively changing from N→64→128→256 channels.

This dynamic normalization approach was complemented by group normalization with 8 groups[24] for the skip connection outputs[25]:

$$h_{\text{combined}} = \text{GroupNorm8}(h_3 + h\text{skip}) \tag{4}$$

and layer normalization in the output heads:

$$h_{\text{norm}} = \text{LayerNorm}(h_{\text{pooled}}) \tag{5}$$

| Category | Parameter | Value/Setting |
|---|---|---|
| **Neural Network Architecture** | | |
| Input Layer | Neurons | 581 |
| | Sequence Length | 32 timesteps |
| CNN | Initial Channels | 64, 128, 256 |
| | Kernel Size | 3 |
| | Skip Connection | 1×1 convolution |
| LSTM | Hidden Size | 128 |
| | Number of Layers | 2 |
| | Direction | Bidirectional |
| Attention | Number of Heads | 8 |
| | Input Dimension | 256 |
| **Training Parameters** | | |
| Optimization | Learning Rate | $1 \times 10^{-4}$ |
| | Weight Decay | $1 \times 10^{-5}$ |
| | Batch Size | 32 |
| | Maximum Epochs | 500 |
| **Regularization** | | |
| Dropout | Deep Layers | 0.5 |
| | CNN Layers | 0.25 |
| Other | Gradient Clipping | 1.0 |
| | Early StoppingPatience | 7 epochs |
| **Loss Function** | | |
| Focal Loss | $\alpha$ | 2.0 |
| | $\gamma$ | 2.0 |
| Task Weights | Multiclass | 1.0 |
| | Left/Right Reach | 0.5 each |
| | Neuronal Activity | 1.0 |
| **Data Split** | | |
| Training | Proportion | 0.7 |
| Validation | Proportion | 0.15 |
| Test | Proportion | 0.15 |

**Table 2.** Hyperparameters and Configuration Details for the Proposed Hybrid Model. The number of neurons input varies across datasets due to difference in the number of neurons for each field of view (FOV). Data 1A (581 neurons), Data 2A (593 neurons), Data 1B (421 neurons), and Data 2B (431 neurons). The model architecture automatically adapts to the input dimension while maintaining all hyperparameters. The values presented in this table correspond to Data 1A (581 neurons).

This multi-level normalization strategy adaptively normalized neuronal activity patterns based on the current batch statistics while maintaining stable training across varying batch sizes through group normalization[24]. The skip connections preserved important relative activity patterns between neurons, while layer normalization in the output heads ensured consistent normalization of the final representations[25]. Together, these normalization techniques, applied at different stages of the network, allowed our model to handle the complex statistics of neuronal data while maintaining the flexibility to learn meaningful representations at multiple scales.

The first major component of our hybrid model was a specialized convolutional neural network (CNN) designed specifically for processing high-dimensional neural activity data[19]. This CNN served as the primary feature extractor, transforming raw neuronal recordings into meaningful representations that captured the complex spatial relationships between different neurons' activity patterns[20]. The architecture's design reflected our deep understanding of how M1 networks at neuronal ensembles level encode behavioral information[22], with each layer progressively building more sophisticated representations of neuronal activity patterns. The mathematical formulation of our three-layer CNN architecture is expressed as:

$$h_1 = \text{Dropout}_{0.25}(\text{ReLU}(\text{BN}(W_1 * x + b_1)))$$
$$h_2 = \text{Dropout}_{0.25}(\text{ReLU}(\text{BN}(W_2 * h_1 + b_2)))$$
$$h_3 = \text{ReLU}(\text{BN}(W_3 * h_2 + b_3))$$

(6)

The first layer of our architecture performed a crucial dimensionality reduction, transforming the input from N (here 581) individual neurons into a 64-dimensional feature space. This dimensionality reduction reflected careful optimization through rigorous experimentation[19], achieving a critical balance that preserved essential neuronal activity patterns while effectively filtering out recording noise and neuronal variability. Through extensive testing and validation, we determined that the 64-dimensional projection space optimally captured the underlying neuronal dynamics while enhancing the signal-to-noise ratio of behaviorally relevant patterns[20]. We used a kernel size of 3 for this layer based on the spatial organization of neuronal circuits, as this size effectively captured local interactions between neighboring neurons in the recording field of view[22]. The operation of this layer began with a convolution that identifies local patterns in neuronal activity, followed by batch normalization to stabilize neuronal activity distributions across different recording sessions. Moving to the second layer, we expanded the feature space from 64 to 128 channels. This expansion was essential because neuronal representations of behavior rarely rely on simple, isolated patterns[22]. Instead, they typically involve coordinated activity across multiple neuronal subpopulations. The increased dimensionality allowed our model to detect and represent these complex interactions between different neuronal assemblies. This layer became particularly important in capturing temporal dependencies in activity patterns and identifying population-level activity signatures that preceded specific behaviors[19]. The third and final convolutional layer further expanded the feature space to 256 channels, creating rich, high-level representations of coordinated neural activity. This expansion provided the capacity needed to represent complex forelimbs movement states through sophisticated combinations of neuronal dynamics[20]. This enabled the detection of subtle variations in population activity that might be crucial for movement prediction and created a feature space with sufficient dimensionality for the subsequent attention mechanism to effectively identify and focus on the most relevant patterns. A critical innovation in our architecture was the implementation of a skip connection using a 1×1 convolution[25], mathematically expressed as:

$$h_{\text{skip}} = W_{\text{skip}} * x, \quad W_{\text{skip}} \in \mathbb{R}^{256 \times 581 \times 1}$$

(7)

The skip connection served as a vital pathway for preserving fine-grained neuronal activity patterns that might otherwise be lost through the main convolutional layers. In fact, provided a direct route for gradient flow during training, which was essential for learning subtle neuronal patterns, crucial for movement prediction. The 1×1 convolution was particularly powerful as it allowed the network to learn optimal combinations of raw neuronal activity while maintaining the original temporal resolution of the recordings.

The output from the skip connection was combined with the main pathway through element-wise addition and normalized usi ng group normalization[24]:

$$h_{\text{combined}} = \text{GroupNorm}_8(h_3 + h_{\text{skip}}) \tag{8}$$

The implementation of group normalization with 8 groups emerged as a superior choice over batch normalization after extensive analysis of our distributed training requirements. This decision was motivated by group normalization's superior stability across varying batch sizes and its consistent performance in normalizing high-dimensional neuronal activity patterns across multiple GPUs. Unlike batch normalization, group normalization maintained reliable statistics even when processing data distributed across different computational nodes, ensuring consistent feature normalization regardless of how neuronal data was partitioned during training.

The progressive increase in channel dimensions (581→64→128→256) throughout our CNN architecture enabled the network to build a hierarchical understanding of neuronal ensemble dynamics. At the local circuit level, the first layer captured correlations between nearby neurons, reflecting the interactions within local neuronal circuits. As we moved to the subpopulation level in the second layer, the network identified patterns within neuronal assemblies that might be involved in specific aspects of movement preparation or execution. Finally, at the population level, the third layer integrated information across the entire recorded population, creating comprehensive representations that can capture complex movement states.

Following CNN's spatial feature extraction, our model employed a BiLSTM network to process the temporal dynamics of neuronal activity patterns. Preparatory neural activity often precedes movement execution by hundreds of milliseconds, while the consequences of movement execution continue to influence neural activity patterns well after the movement is initiated[14, 26]. Since neuronal encoding of movement involved complex temporal sequences that unfold both before and during the execution of complex movement, a BiLSTM network was used to capture complex temporal relationships in neuronal data by processing neuronal sequences in both forward and backward directions:

$$\overrightarrow{h_t} = \text{LSTM}_{\text{forward}}(x_t, \overrightarrow{h_{t-1}}, \overrightarrow{C_{t-1}}) \\ \overleftarrow{h_t} = \text{LSTM}_{\text{backward}}(x_t, \overleftarrow{h_{t+1}}, \overleftarrow{C_{t+1}}) \tag{9}$$

The forward direction ($\overrightarrow{h_t}$) processed neuronal activity patterns as they unfold in time, capturing how neuronal population dynamics evolve leading up to a complex movement of footstep. This was crucial for identifying preparatory neuronal activity that might predict upcoming behaviors. The backward direction ($\overleftarrow{h_t}$) processed the sequence in reverse, allowing the model to understand how future events influence the interpretation of current neuronal states[20]. This bidirectional processing was important because motor preparation often involves subtle changes in neuronal activity that become meaningful only in the context of subsequent movements. The LSTM cell computations for each direction followed the standard equations[14], but with specific adaptations for neuronal data processing:

$$f_t = \sigma(W_f \cdot [h_{t-1}, x_t] + b_f) \quad \text{(forget gate)} \tag{10}$$

The forget gate $f_t$ determined which aspects of the previous neuronal activity dynamics should be retained or discarded[14]. In our context, $W_f$ learned to identify which features of past neuronal states were predictive of upcoming movements[19]. This was crucial because different neurons exhibit varying temporal persistence in their activity patterns[22]. For instance, some neurons might maintain elevated activity rates throughout the entire preparation phase of a complex movement, while others might show transient activation. The forget gate learned these neuron-specific temporal characteristics. The sigmoid activation function σ outputs valued between 0 and 1, effectively creating a filter that could selectively maintain or eliminate different aspects of the neuronal activity profile.

$$i_t = \sigma(W_i \cdot [h_{t-1}, x_t] + b_i) \quad \text{(input gate)} \tag{11}$$

The input gate $i_t$ controlled how new neural activity patterns were incorporated into the cell's memory[21]. The weight matrix $W_i$ learned to identify significant changes in neuronal activity that could be behaviorally relevant[22]. This was essential because not all changes in neuronal activity rates were meaningful for movement prediction. For example, when a mouse begun preparing for a footstep, certain neuronal populations might show subtle but important changes in their activity patterns. The input gate learned to recognize these behaviorally relevant changes while being less sensitive to random fluctuations in neuronal activity.

$$\tilde{C}t = \tanh(W_C \cdot [ht - 1, x_t] + b_C) \quad \text{(candidate cell state)} \tag{12}$$

The candidate cell state $\tilde{C_t}$ proposed new information to be added to the cell's memory based on the current neuronal activity pattern $x_t$ and the previous hidden state $h_{t-1}$[21]. The hyperbolic tangent activation function tanh squashed the values between -1 and 1, which was particularly suitable for neuronal data as it could represent both increases and decreases in neuronal signal rates relative to baseline activity[19]. The weight matrix $W_C$ learned to transform current neuronal activity patterns into a format that could be effectively combined with the existing cell state.

$$C_t = f_t * C_{t-1} + i_t * \tilde{C}_t \quad \text{(cell state update)} \tag{13}$$

The cell state update equation combined the selective forgetting of old information ($f_t * C_{t-1}$) with the selective addition of new information ($i_t * \tilde{C}_t$). This mechanism was important for tracking the evolution of neuronal population activity over time. The cell state could maintain a memory of sustained neuronal activity patterns (like preparatory activity) while still being able to update when significant changes occur (like the initiation of footstep). The element-wise multiplication operations (*) allowed different elements of the cell state to be updated independently, which was essential because different neuronal su bpopulations might change their activity patterns at different times.

$$o_t = \sigma(W_o \cdot [h_{t-1}, x_t] + b_o) \quad \text{(output gate)} \tag{14}$$

The output gate $o_t$ controlled what information from the cell state should be exposed to the next layer of the network. In our neural processing context, $W_o$ learned to identify which aspects of the maintained neuronal activity patterns were most relevant for movement prediction at each time step, acknowledging the relevance of different neuronal patterns depending on the phase of movement (preparation, initiation, or execution).

$$h_t = o_t * \tanh(C_t) \quad \text{(hidden state)} \tag{15}$$

The hidden state $h_t$ represented the final output of the LSTM cell at each time step. The hyperbolic tangent of the cell state $\tanh(C_t)$ was filtered by the output gate $o_t$ through element-wise multiplication. This created a refined representation of the neuronal activity that emphasized behaviorally relevant patterns while suppressing less important variations. The hidden state captured both the immediate neuronal activity pattern and its temporal context, which was essential for predicting upcoming movements.

The architecture's hyperparameters were carefully tuned based on the characteristics of neuronal population dynamics. The hidden state dimension of 128 was chosen to provide sufficient capacity for representing complex temporal patterns while maintaining computational efficiency. This dimensionality allowed the network to capture both broad population-level dynamics and subtle temporal variations in neuronal activity that might be predictive of movement.

The implementation used two BiLSTM layers with a dropout rate of 0.5 between them. This two-layer architecture created a hierarchical representation of temporal patterns; the first layer captures basic temporal dependencies in neuronal activity patterns, while the second layer learned more abstract temporal relationships that might span longer time intervals. The relatively high dropout rate of 0.5 was chosen based on empirical testing and served to prevent the model from overfitting specific temporal patterns, ensuring robust generalization across different movement sessions. The forget gate ($f_t$) played a particularly crucial role in our context, as it allowed the network to selectively maintain or discard information about past neuronal states. This was essential because different neurons may maintain their activity states for varying durations, and the relevance of past neuronal activity could vary depending on the behavioral context. The input gate ($i_t$) controlled how new neuronal activity patterns were incorporated into the cell state, while the output gate ($o_t$) determined how the internal cell state influences the hidden state representation. The cell state ($C_t$) served as a memory mechanism that could maintain information about neuronal activity patterns over extended time periods. This was important for capturing slow-evolving preparatory activity that might begin several hundred milliseconds before a reaching movement. The candidate cell state ($\tilde{C}_t$) proposed updates to this memory based on current neuronal activity patterns and the previous hidden state. The combination of forward and backward processing created a rich representation of temporal context for each time point in the neuronal sequence. For any given moment, the model had access to both the history of activity leading up to that moment and information about future neuronal states. This bidirectional context was crucial for accurate movement prediction allowing the model to distinguish between similar neuronal activity patterns that may lead to different behavioral outcomes based on their broader temporal context.

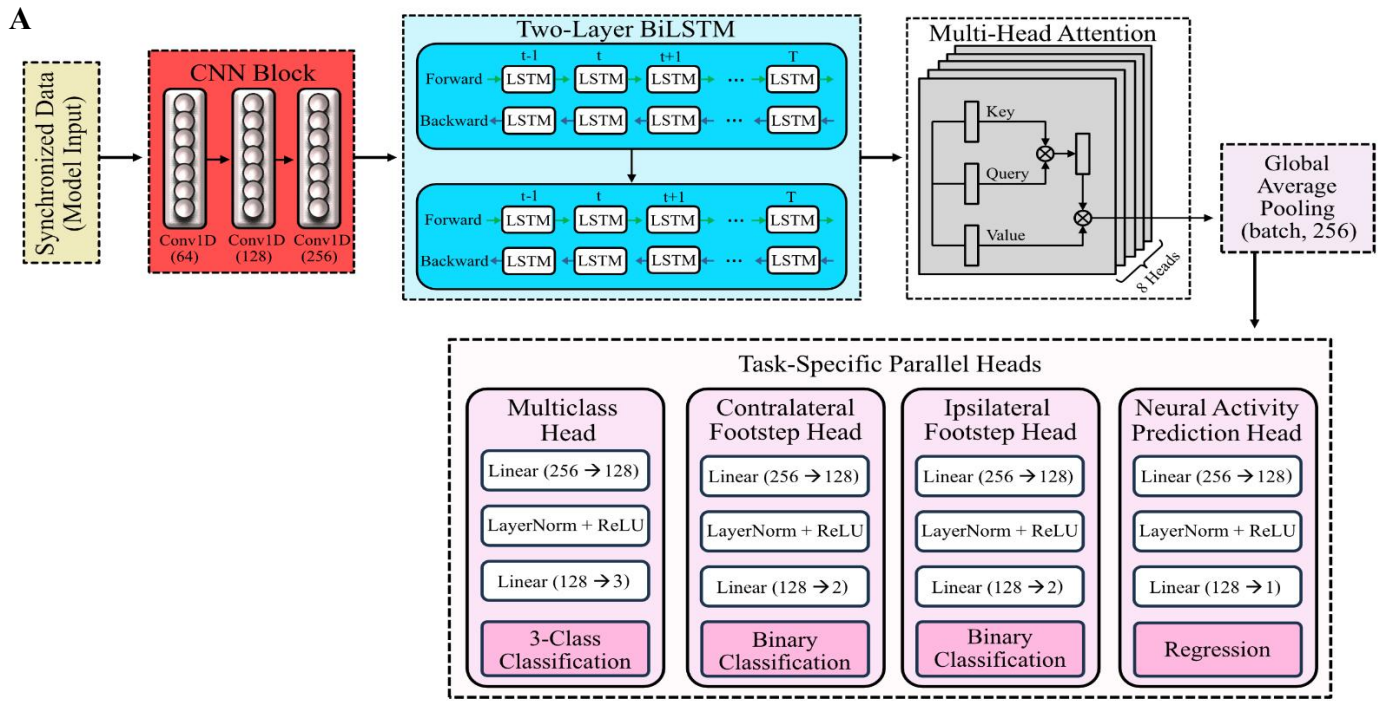

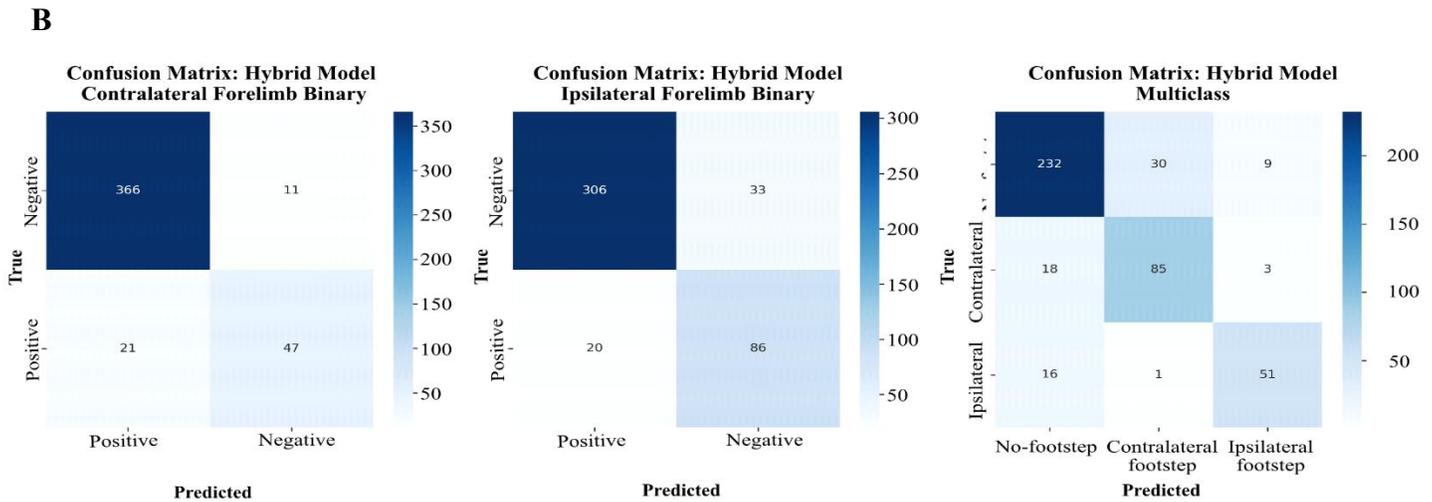

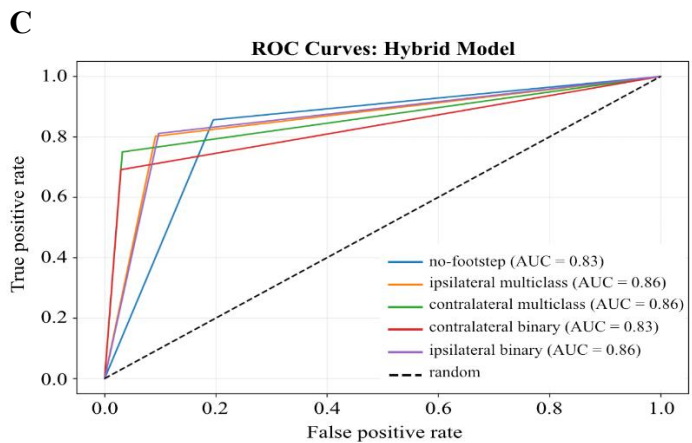

**Figure 3. Architecture and performance of attention-based hybrid CNN-BiLSTM model in decoding bilateral forelimb complex movements from unilateral M1. A)** Illustration demonstrates the architecture of

the hybrid model. **B)** Confusion matrices for hybrid model in decoding forelimbs footsteps on the grid-walking wheel. The multiclass confusion matrix (right) highlights strong accuracy in classifying no-footstep states (232 correct) and distinguishing ipsilateral (85 correct) and contralateral footsteps (51 correct). The binary classifiers (center and left) demonstrate improved discrimination for single-forelimb detection, with contra- and ipsilateral footstep classifiers achieving 366/47 and 306/86 true negatives/positives, respectively. The balanced performance across forelimbs suggests symmetric neuronal encoding, while false positives may indicate detected preparatory activity without the execution of movement. These results validate the model's ability to extract meaningful bilateral forelimb motor patterns from unilateral M1 neuronal activity. **C)** Represents ROC curves comparing the hybrid model's performance in multiclass and binary classification of forelimbs footsteps. The plot includes five curves: three from the multiclass classifier (no-footstep, contralateral forelimb footstep, and ipsilateral forelimb footstep) and two from binary classifiers (contralateral and ipsilateral forelimbs footsteps), evaluated against random chance (diagonal dashed line). All classifiers demonstrate strong discriminative ability with AUC values ranging from 0.83 to 0.86, where ipsilateral detection (AUC = 0.86) slightly outperforms contralateral detection (AUC = 0.83), indicating robust performance substantially above random classification (AUC = 0.5).

It also helped the model understand how different phases of movement (preparation, initiation, execution) were reflected in the temporal evolution of neuronal ensemble activity. Following the temporal processing by our BiLSTM layers, a multi-head self-attention [8] mechanism was implemented to identify and weigh the relative importance of different temporal patterns in neuronal activity. Each attention head could specialize in detecting distinct aspects of neuronal patterns, creating a comprehensive representation of behaviorally relevant neuronal dynamics[21].

The attention computation for each head operated through a query-key-value mechanism, mathematically expressed as[21]:

$$\text{head}_i = \text{Attention}(XW_i^Q, XW_i^K, XW_i^V) \tag{16}$$

In this formulation, the input neural sequence X was projected into three different spaces through learned weight matrices $W_i^Q$, $W_i^K$, and $W_i^V$. These projections served distinct purposes in our neural processing context: the query projection (XWiQ) represented the current neuronal state, the key projection (XWiK) helped identify relevant patterns in the sequence, and the value projection (XWiV) containing the actual neuronal information to be extracted. The attention computation then proceeded as[21]:

$$\text{Attention}(Q, K, V) = \text{softmax}\left(\frac{QK^T}{\sqrt{d_k}}\right)V \tag{17}$$

The scaling factor dk played a crucial role in maintaining stable gradients during training, particularly important given the high dimensionality of our neuronal data. The softmax operation created a probability distribution over temporal positions, effectively determining which time points in the neuronal sequence contained relevant information for movement prediction.

The multi-head configuration allowed parallel processing of neuronal patterns at different temporal scales:

$$\text{MultiHead}(X) = \text{Concat}(\text{head}_1, ..., \text{head}_8)W^O \tag{18}$$

Each head could specialize in different aspects of activity. For instance, some heads might focus on rapid changes in firing rates that signal movement initiation, while others might track slower, sustained changes associated with movement preparation. The output projection matrix $W_O$ combined these different perspectives into a unified representation.

This attention mechanism proved particularly powerful in our neuronal processing context for several reasons. First, it could dynamically adjust its focus based on the current behavioral context, giving more weight to neuronal patterns that are predictive of upcoming movements. Second, the parallel processing of multiple attention heads allowed simultaneous tracking of different neuronal assemblies that might become active at different times during movement preparation and execution. Third, attention weights provided interpretable insights into which neuronal activity patterns the model considered most relevant for movement prediction.

The final component of our hybrid model comprised three specialized output heads that work in synergy during both training and inference. Each head processed the shared neural representations from our attention mechanism to extract complementary aspects of the movement predictions. During training, these heads learned simultaneously[19], allowing the model to discover neuronal activity patterns that were informative across multiple prediction tasks. The multiclass classification head formed the primary behavioral predictor, distinguishing between three distinct states: no footstep, ipsilateral forelimb footstep, and contralateral forelimb footstep. We implemented this through a structured pipeline of layer normalization[23], ReLU activation, and dropout regularization, followed by a softmax operation that produced interpretable probabilities for each state. This architecture ensured stable neuronal representations across recording sessions while preventing over-reliance on specific neuronal patterns through a dropout rate of 0.5. While the multiclass head provided a complete categorical prediction, we additionally employed separate binary classification heads for ipsi- and contralateral forelimbs movements. This parallel prediction strategy allowed the model to develop specialized feature detectors for each forelimb's movements[20]. The binary heads could detect weak movement signals that might be missed when forcing a multiclass classification[19]. Moreover, the agreement or disagreement between binary and multiclass predictions provided a measure of prediction confidence, particularly valuable in real-time applications. The neuronal activity prediction head completed our multi-task architecture by predicting future neuronal states through a linear output layer. This head served multiple crucial functions beyond simple prediction. As a regularizer, it constrained the model's internal representations to capture the underlying dynamics of neuronal population activity, preventing overfitting to purely categorical features. The prediction of continuous neuronal activity forced the model to learn more nuanced representations that reflected the graded nature of neuronal coding. Additionally, this head helped maintain sensitivity to subtle changes in neuronal activity patterns that anticipate movement execution. This multi-task design created a comprehensive prediction framework where each head contributed unique strengths while supporting the others.

*Loss Function Design and Training Strategy*

Hybrid neural network training relied on a multitask loss function that addressed the unique challenges of neuronal data processing and movement prediction. The loss function combined multiple components, each serving a specific purpose in guiding the model's learning process:

$$\mathcal{L}_{\text{total}} = w_1 \mathcal{L}_{\text{focal}}^{\text{multi}} + w_2 \mathcal{L}_{\text{focal}}^{\text{left}} + w_3 \mathcal{L}_{\text{focal}}^{\text{right}} + w_4 \mathcal{L}_{\text{MSE}}^{\text{neural}} \tag{19}$$

where $L^{multi}_{focal}$ represented the loss of multiclass classification, $L^{left}_{focal}$ and $L^{right}_{focal}$ were binary classification losses for each forelimb, and $L^{neural}_{MSE}$ was the mean squared error for the prediction of neuronal activity. The weights $w_1, w_2, w_3$, and $w_4$ balanced these components' contributions during training.

Our classification tasks faced a fundamental challenge in neuronal behavioral data: the substantial imbalance between movement (footstep) and non-movement (non-footstep) periods. In our recordings, footsteps occurred sparsely, while periods of non-footstep dominated the dataset. To address this imbalance, we implemented focal loss, an advanced adaptation of cross-entropy that automatically adjusted the learning process to emphasize rare but important movement events[27]. Unlike standard cross-entropy, which treated all classification errors equally, focal loss dynamically modulated the contribution of each example based on the model's current performance. The mathematical formulation of focal loss captured this adaptive behavior:

$$\mathcal{L}_{focal} = -\alpha(1 - p_t)^\gamma \log(p_t) \qquad (20)$$

where $p_t$ represented the model's estimated probability for the true class. Through empirical experimentation, we determined optimal values for the focal loss parameters: α = 2.0 and γ = 2.0. The α parameter adjusted the weight of rare classes (footstep events) relative to common classes (non-footstep), effectively balancing their contributions to the gradient updates. The γ parameter modulated the rate at which well-classified examples were down-weighted, allowing the model to focus on challenging cases where neuronal patterns were more ambiguous.

For neuronal activity prediction, we employed mean squared error loss, which effectively captured the continuous nature of calcium signaling rates (neuronal activity). This combination of focal loss for discrete behaviors and MSE for continuous neuronal activity created a balanced learning objective that respects both the categorical and continuous aspects of the prediction tasks[27].

The training of the hybrid neural network presented unique challenges due to the complex nature of neuronal networks data, including high variability in calcium signaling patterns (calcium firing rates) indicating neuronal activity, temporal dependencies, and the need to capture both fine-grained neuronal patterns and broader behavioral states. To address these challenges, we implemented a training strategy that combines several optimization techniques. At the core of our optimization approach lied the AdamW optimizer[28], an enhanced variant of the Adam optimizer that implemented a corrected weight decay regularization:

$$\theta_{t+1} = \theta_t - \eta_t (\nabla f_t(\theta_t) + \lambda \theta_t) \qquad (21)$$

In this formulation, θt represented ther model parameters at time step t, ∇ft(θt) was the gradient of loss function, and λθt implemented weight decay regularization. The inclusion of weight decay with λ = 1e−5 provided essential regularization by penalizing large weight values, helping the model to prevent from learning spurious patterns in the variable neuronal data. A critical component of our training strategy was the learning rate schedule, which followed a one-cycle pattern:

$$\eta_t = \eta_{max} \cdot \min\left(\frac{t}{T_{warmup}}, 1 - \frac{t - T_{warmup}}{T_{total} - T_{warmup}}\right) \qquad (22)$$

We set the maximum learning rate $\eta_{max}$ to 1e−4 after careful empirical validation. This schedule incorporated a warmup period ($T_{warmup} = 0.3 \cdot T_{total}$) during which the learning rate gradually increased, allowing the model to establish stable gradient statistics before reaching full learning rates. This approach was particularly important for neural data, where initial gradient estimates could be noisy due to the high variability in neuronal activity patterns. To ensure training stability and prevent the common challenges associated with deep neural networks processing temporal data, we implemented several crucial safeguards which are elaborated. Gradient clipping enforced a maximum norm of 1.0 on parameter gradients. This constraint was essential when working with neuronal sequence data, as it prevented gradient explosions that could occur when backpropagating through long temporal dependencies in neuronal activity patterns.

An early stopping mechanism was also employed to monitor validation loss with a patience of 7 epochs[29]. This mechanism prevented overfitting by stopping training when the model's performance on the validation set stopped improving, while still allowing sufficient time for the model to learn meaningful representations of neuronal activity patterns. Layer normalization was also applied after each major component in our architecture[23]. This helped maintain stable neuronal activations throughout the network, given the varying scales and distributions of calcium firing rates across different neurons and time periods. Dropout regularization with a probability of 0.5 was implemented in deeper layers, forcing the model to learn redundant representations and preventing over-reliance on specific neuronal activity patterns[19].

This comprehensive training strategy created a robust framework for learning from complex neuronal networks data extracted from *in vivo* 2P GCaMP6s imaging while maintaining good generalization capabilities. The combination of adaptive learning rates, careful regularization, and stability measures allowed the model to capture both the fine temporal dynamics of neuronal activity and the broader patterns that predicted behavioral states.

**Hybrid model enhances performance in decoding skilled bilateral forelimb movement**

The comparative analysis in Table 3 demonstrates the superior performance of the hybrid CNN-BiLSTM architecture over the traditional LSTM model across all performance metrics achieved from all datasets. Prediction of contralateral footstep achieved the highest accuracy of 93.10% from hybrid model compared to 81.10% accuracy for LSTM model (dataset 1B, table 3). The metrics for ipsilateral prediction of same dataset was 90.07% (82.20% for LSTM). The performance improvements were even more remarkable for multiclass classification accuracy, with a consistent performance boost across datasets.

| Dataset | Model | Task | Accuracy | Precision | Recall | F1-score |
| --- | --- | --- | --- | --- | --- | --- |
| 1A | LSTM | Contralateral | 0.822 | 0.601 | 0.737 | 0.632 |
| | | Multiclass | 0.656 | 0.723 | 0.653 | 0.646 |
| | | Ipsilateral | 0.776 | 0.554 | 0.848 | 0.703 |
| | Hybrid | Contralateral | 0.911 | 0.857 | 0.810 | 0.833 |
| | | Multiclass | 0.842 | 0.842 | 0.842 | 0.841 |
| | | Ipsilateral | 0.906 | 0.799 | 0.810 | 0.804 |
| 2A | LSTM | Contralateral | 0.817 | 0.765 | 0.789 | 0.776 |

|  |  |  | Multiclass | 0.731 | 0.753 | 0.731 | 0.732 |
|---|---|---|---|---|---|---|---|
|  |  |  | Ipsilateral | 0.815 | 0.682 | 0.750 | 0.714 |
|  |  | Hybrid | Contralateral | 0.909 | 0.819 | 0.819 | 0.819 |
|  |  |  | Multiclass | 0.835 | 0.835 | 0.835 | 0.835 |
|  |  |  | Ipsilateral | 0.907 | 0.821 | 0.855 | 0.838 |
| 1B | LSTM | | Contralateral | 0.811 | 0.604 | 0.756 | 0.671 |
|  |  |  | Multiclass | 0.629 | 0.740 | 0.629 | 0.627 |
|  |  |  | Ipsilateral | 0.822 | 0.500 | 0.425 | 0.459 |
|  | Hybrid | | Contralateral | 0.931 | 0.792 | 0.784 | 0.788 |
|  |  |  | Multiclass | 0.847 | 0.846 | 0.847 | 0.846 |
|  |  |  | Ipsilateral | 0.901 | 0.831 | 0.745 | 0.786 |
| 2B | LSTM | | Contralateral | 0.844 | 0.717 | 0.761 | 0.738 |
|  |  |  | Multiclass | 0.682 | 0.711 | 0.682 | 0.685 |
|  |  |  | Ipsilateral | 0.827 | 0.577 | 0.460 | 0.512 |
|  | Hybrid | | Left Foot | 0.918 | 0.769 | 0.816 | 0.792 |
|  |  |  | Multiclass | 0.849 | 0.849 | 0.849 | 0.848 |
|  |  |  | Ipsilateral | 0.906 | 0.876 | 0.809 | 0.841 |

**Table 3**. Comparative analysis of performance prediction for two proposed deep learning models. Performance was evaluated across datasets (1A & 2A mouse A, and 1B & 2B mouse B). Tasks were categorized into ipsilateral forelimb, contralateral forelimb, and multiclass classifications.

The hybrid model achieved improvement of 85.00% accuracy in multiclass prediction compared to the LSTM's 62.9%, representing a substantial enhancement in the model's ability to simultaneously distinguish between no footstep, ipsilateral, and contralateral footsteps. The confusion matrices revealed the hybrid model's superior discrimination capabilities, with significantly reduced false positives and false negatives compared to the LSTM model (Fig.3). Quantitatively, the hybrid model achieved 366/47 and 306/86 true negatives/positives for contralateral and ipsilateral binary classification respectively, with minimal false positives (11 and 33) and false negatives (21 and 20). In multiclass prediction, the model correctly identified 232 no-footstep states, 85 ipsilateral, and 51 contralateral footsteps. This improvement was further illustrated by the ROC curves, which demonstrate the hybrid model's enhanced ability to maintain high true positive rates while minimizing false positives across different classification thresholds, with AUC values ranging from 0.83 to 0.86, substantially above random classification (AUC = 0.5). These results provided compelling evidence that our hybrid architecture's integration of CNN-based spatial feature extraction with BiLSTM temporal processing created a more powerful framework for decoding complex bilateral forelimb movements from unilateral M1 neuronal networks activity achieved from *in vivo* 2P calcium imaging. The consistent performance improvements across all metrics and datasets validated our architectural design choices, demonstrating that the combination of convolutional processing for spatial features and BiLSTM for temporal dependencies significantly enhanced the model's ability to capture the intricate patterns in neuronal networks activity associated with complex movement execution.

## DISCUSSION

Decoding behavior, such as complex movements, from multiscale brain networks relies on complexity of extracted signals and accuracy of the encoder. The optimized complex signals should encompass both spatial and temporal information of a monitored brain network. We used advanced *in vivo* 2P calcium imaging on the head-fixed grid-walking mouse to extract multi-dimensional sets of data from M1 excitatory neuronal networks. This approach enabled the monitoring of individual neuron activity, integrating it with neuronal ensemble dynamics and complex motor movement (gait). To decode motor movement, we first implemented a supervised fully connected layers LSTM and then proposed a hybrid model that captured both spatial and temporal characteristics of 2P imaging data for multiclass predictions. By implementing CNN, this approach could learn the spatial features of neurons in a captured neuronal network, while a BiLSTM with an attention mechanism could detect complex temporal dependencies across neuronal events in time series. We demonstrated that ipsi- and contralateral forelimbs complex motor movements can be decoded from neuronal signals achieved from 2P calcium imaging of neuronal networks in unilateral M1 using supervised deep learning models. This suggests that layer II/III excitatory neuronal ensembles in M1 of one hemisphere could provide essential information for execution of a complex movement in both forelimbs. Moreover, the improved performance metrics achieved by our hybrid model (compared to LSTM) suggests the existence of hidden classes of information within the spatial distribution of neurons in M1 networks, highlighting their crucial role in behavior decoding. The motor task evaluated here consisted of both locomotion like and reaching-grasping movements representing complex gait movement[12, 28, 30]. Several brain injuries such as stroke or TBI result in gait deficits[16, 31]. The ability to decode bilateral forelimbs movements from neuronal networks of a single hemisphere, at the level of single neuron and neuronal ensemble, can be crucial in various applications, such as building an encoding platform for neural prosthesis for individuals with brain injuries and paralysis, or for complex BMIs controlling additional limbs.

In comparison with previous studies that used deep machine learning models to decode movement based on 2P calcium imaging data, our approach presents several advantages and novelties[32, 33]. First, our behavioral paradigm represents a complex movement that requires skilled motor behavior. Second, to our knowledge, we are among the first to employ a hybrid CNN-BiLSTM model, incorporating attention mechanisms and focal loss to classify skilled forelimbs movement from *in vivo* 2P calcium imaging extracted signals at level of individual neurons and neuronal ensembles. The model's ability to capture both spatial and temporal dependencies from neuronal network data, combined with its interpretability through permutation importance analysis, demonstrated the effectiveness of integrating convolutional and recurrent neural networks for analyzing complex neuronal network dynamics in unilateral M1 neuronal ensembles and their relationship to intricate bilateral forelimbs motor behavior. Finally, this study establishes a new benchmark for application of hybrid deep learning models in decoding behavior from advanced *in vivo* 2P imaging data.

## METHODS

*Experimental model*

All procedures were performed in accordance with the National Institute of Health Animal Protection Guidelines. CaMKII-tTA/tetO-GCaMP6s transgenic mice were achieved from crossing B6;DBA-Tg(tetO-GCaMP6s)2Niell/J (The Jackson Laboratory) with B6.Cg-Tg(CamK2a-tTA)1Mmay/DboJ (The Jackson Laboratory) in order to have constant expression of GCaMP6s. 4 months old CaMKII-

tTA/tetO- GCaMP6s males were used in this study. Mice were housed under pathogen-free conditions and were maintained on a 12h light/dark cycle with free access to food and water.

*General surgery*

Each mouse underwent a single surgery. Surgical procedures were performed under 1% isoflurane anesthesia with initial induction at 4%. Craniotomy was performed in a stereotactic frame (David Kopf Instruments, CA, USA) using a hand-held dentist drill with 0.1 mm burr. The scalp was shaved, cleaned, and resected, the skull was cleaned, and the wound margins glued to the skull with tissue glue (VetBond, 3M), and a 4 mm circular craniotomy was made over the forelimb motor cortical region (left hemisphere) to make the possibility of recording M1 neuronal networks. An 5mm round coverslips (Harvard Apparatus) was then placed on the craniotomy and glued in place first with tissue glue (VetBond, 3M). The precis location of bregma was then labeled on the coverslips detecting and stabilizing the FOVs. A costume-made stainless-steel head-bar was glued (Krazy Glue) to the neck of each animal for head-fixed recording. The skull surface, the glass plate, and the base of the head-bar were then covered with dental acrylic powder (Ortho Jet; Lang Dental). Mice were then left for three weeks to recover.

*Behavioral training and Forelimb tracking*

Mice were handled extensively before being head restrained and habituated to the grid- walking wheel. A costume-made grid walking wheel was created to track complex movement under the 2P imaging. A 1 x 1 cm squares grid was cut into 7-inch-wide sections and fit into the backs of a 9- inch-wide section. To motorize the wheel, a small DC motor (Greartisan DC12V, RPM5) was connected to the axle of the wheel such that the grid could be driven forward. This adaptation forces the mouse to move forward with adjustable velocity. Behavior was recorded using an infrared-sensitive camera (Genie Nano, Teledyne DALSA) that was mounted facing the mouse. Infrared illuminations enabled behavioral tracking while performing 2P microscopy in a darkened microscope enclosure. After the period of habituation and training, all sessions were recorded at the same velocity of the wheel. Behavioral analysis was performed manually based on previously established criteria and footsteps frames were labeled[12, 16].

*2P calcium imaging and analysis*

*In vivo* 2P photon calcium imaging was performed using an Ultima 2Pplus microscope (Bruker 453 Corporation) controlled by Prairie View software (Bruker Corporation). The GCaMP6s excitation source for imaging was a fixed wavelength 920 nm femtosecond laser (Axon 920 from Coherent Inc.). A 16×/0.8 NA (Nikon) water immersion objective was used, yielding FOV size of 1135.5 x 1135.5 µm. Resonant galvo scanner was used to acquire images at 15 Hz (15 frames per second). Raw recorded files were converted to tiff files using Prairie View software. Custom-written MATLAB software and ImageJ (NIH) were used to analyze extracted data[9, 12]. Regions of interest (ROIs) related to somas with an intensity higher than 30% of the background intensity (darkest region of a video) were selected for each FOV, as described previously. These ROIs were also size constrained from 40 to 120 µm$^2$. The output matrix was achieved with all potential ROIs and their fluorescent traces, and each ROI was manually inspected for its morphology and fluorescent traces. Fluorescence intensity was converted to $dF/F_0$ values, which were calculated by subtracting the average fluorescence of the ROI from the current fluorescence at each frame, then dividing by the average fluorescence ($F(t) – F_0)/F_0$). Calcium transient peaks were detected by applying a MATLAB smoothing and peak detection function to the waveform. Peaks that were not greater than the root mean square of $dF/F_0$ were not considered calcium transients. A deconvolution algorithm was then applied to all dF/F$_0$ traces to remove the non-physiological slow decay of GcaMP6s signals and to sharpen the calcium transients[17].

*Training and evaluation of LSTM-FCNN model*

The LSTM model was implemented with standardized neuronal activity input from identified number neurons (e.g., 581 neurons), processed through a carefully designed sequence-based structure with a length of 10 timesteps to capture temporal dependencies in neural activity patterns. This temporal window was chosen to balance between capturing sufficient movement-related neural dynamics and maintaining computational efficiency. The core of our architecture consisted of a two-layer LSTM network with 64 hidden units in each layer, incorporating dropout mechanisms with 0.5 probability between layers to prevent overfitting (Table 1). Following the LSTM layers, we implemented batch normalization to stabilize the training process and normalize the neuronal activity representations. The model then processed these representations through a series of fully connected layers, starting with a reduction to 64 dimensions, followed by a further reduction to 32 dimensions, before reaching the final classification layer. Each of these fully connected layers employed ReLU activation functions and includes dropout layers to ensure robust generalization.

To address the inherent class imbalance in our behavioral data, where movement events (footsteps) were naturally less frequent than non-movement (no footstep) periods, we implemented a dynamic class weighting system. This system automatically calculates appropriate weights based on class frequencies in the training data, ensuring the model pays sufficient attention to the less frequent footstep events while maintaining balanced learning across all behavioral states. The weights were computed using the inverse frequency of each class, scaled by the total number of samples, which effectively compensates for the uneven distribution of footstep events in our recordings. The training process incorporated several technical innovations for robust learning. A learning rate scheduler monitored validation performance and reduced the learning rate by half when performance plateaued, allowing for fine-tuned optimization. Early stopping with a patience of 7 epoch prevented overfitting by halting training when validation performance stops improving. We also employed gradient clipping with a maximum norm of 1.0 to ensure stable training, particularly important when dealing with temporal sequences of neuronal activity. Additionally, L2 regularization with a weight decay of 1e-5 provided further protection against overfitting (Table 1).

## Acknowledgements

This study was supported by the National Science Foundation (NSF grant 2242771) and GM109098-09S1 Supplement from NIH Centers of Biomedical Research Excellence (COBRE).

## Author contribution

S.L. conceived and designed the study. J.C. analyzed behavioral data. G.M. and S.L. conducted experiments, designed the software, and analyzed the data. G.M. and S.L. wrote the manuscript.

## Data and code availability

All data (including raw data, videos of 2P calcium imaging, and behavioral recordings) will be provided upon request. All codes for this study can be found on LatifiLab GitHub (https://github.com/LatifiLab/neural_decoding).

## Competing interest

The authors declare no competing interests.

## References

(1) Betzel, R. F.; Medaglia, J. D.; Bassett, D. S. Diversity of meso-scale architecture in human and non-human connectomes. *Nat Commun* **2018**, *9* (1), 346. DOI: 10.1038/s41467-017-02681-z From NLM Medline.
(2) Latifi, S.; Carmichael, S. T. The emergence of multiscale connectomics-based approaches in stroke recovery. *Trends Neurosci* **2024**, *47* (4), 303-318. DOI: 10.1016/j.tins.2024.01.003 From NLM Medline.
(3) Latifi, S.; DeVries, A. C. Window into the Brain: In Vivo Multiphoton Imaging. *ACS Photonics* **2025**, *12* (1), 1-15. DOI: 10.1021/acsphotonics.4c00958 From NLM PubMed-not-MEDLINE.
(4) Ghanayim, A.; Benisty, H.; Cohen Rimon, A.; Schwartz, S.; Dabdoob, S.; Lifshitz, S.; Talmon, R.; Schiller, J. VTA projections to M1 are essential for reorganization of layer 2-3 network dynamics underlying motor learning. *Nat Commun* **2025**, *16* (1), 200. DOI: 10.1038/s41467-024-55317-4 From NLM Medline.
(5) Inoue, Y.; Mao, H.; Suway, S. B.; Orellana, J.; Schwartz, A. B. Decoding arm speed during reaching. *Nat Commun* **2018**, *9* (1), 5243. DOI: 10.1038/s41467-018-07647-3 From NLM Medline.
(6) Kondapavulur, S.; Lemke, S. M.; Darevsky, D.; Guo, L.; Khanna, P.; Ganguly, K. Transition from predictable to variable motor cortex and striatal ensemble patterning during behavioral exploration. *Nat Commun* **2022**, *13* (1), 2450. DOI: 10.1038/s41467-022-30069-1 From NLM Medline.
(7) Morandell, K.; Huber, D. The role of forelimb motor cortex areas in goal directed action in mice. *Sci Rep* **2017**, *7* (1), 15759. DOI: 10.1038/s41598-017-15835-2 From NLM Medline.
(8) Economo, M. N.; Viswanathan, S.; Tasic, B.; Bas, E.; Winnubst, J.; Menon, V.; Graybuck, L. T.; Nguyen, T. N.; Smith, K. A.; Yao, Z.; et al. Distinct descending motor cortex pathways and their roles in movement. *Nature* **2018**, *563* (7729), 79-84. DOI: 10.1038/s41586-018-0642-9 From NLM Medline.
(9) Latifi, S.; Mitchell, S.; Habibey, R.; Hosseini, F.; Donzis, E.; Estrada-Sanchez, A. M.; Nejad, H. R.; Levine, M.; Golshani, P.; Carmichael, S. T. Neuronal Network Topology Indicates Distinct Recovery Processes after Stroke. *Cereb Cortex* **2020**, *30* (12), 6363-6375. DOI: 10.1093/cercor/bhaa191 From NLM Medline.
(10) Heming, E. A.; Cross, K. P.; Takei, T.; Cook, D. J.; Scott, S. H. Independent representations of ipsilateral and contralateral limbs in primary motor cortex. *Elife* **2019**, *8*. DOI: 10.7554/eLife.48190 From NLM Medline.
(11) Ames, K. C.; Churchland, M. M. Motor cortex signals for each arm are mixed across hemispheres and neurons yet partitioned within the population response. *Elife* **2019**, *8*. DOI: 10.7554/eLife.46159 From NLM Medline.
(12) Latifi, S.; Chang, J.; Pedram, M.; Latifikhereshki, R.; Thomas Carmichael, S. Supervised deep machine learning models predict forelimb movement from excitatory neuronal ensembles and suggest distinct pattern of activity in CFA and RFA networks. *bioRxiv* **2024**, 2024.2001.2030.577967. DOI: 10.1101/2024.01.30.577967.
(13) Livezey, J. A.; Glaser, J. I. Deep learning approaches for neural decoding across architectures and recording modalities. *Brief Bioinform* **2021**, *22* (2), 1577-1591. DOI: 10.1093/bib/bbaa355 From NLM Medline.
(14) Gers, F. A.; Schmidhuber, J.; Cummins, F. Learning to forget: Continual prediction with LSTM. *Neural Comput* **2000**, *12* (10), 2451-2471. DOI: Doi 10.1162/089976600300015015.

(15) Ay, M.; Kulluk, S.; Özbakir, L.; Gülmez, B.; Öztürk, G.; Özer, S. CNN-LSTM and clustering-based spatial-temporal demand forecasting for on-demand ride services. *Neural Comput Appl* **2022**, *34* (24), 22071-22086. DOI: 10.1007/s00521-022-07681-9.

(16) Bechay, K. R.; Abduljawad, N.; Latifi, S.; Suzuki, K.; Iwashita, H.; Carmichael, S. T. PDE2A Inhibition Enhances Axonal Sprouting, Functional Connectivity, and Recovery after Stroke. *J Neurosci* **2022**, *42* (44), 8225-8236. DOI: 10.1523/JNEUROSCI.0730-22.2022 From NLM Medline.

(17) Pnevmatikakis, E. A.; Giovannucci, A. NoRMCorre: An online algorithm for piecewise rigid motion correction of calcium imaging data. *J Neurosci Methods* **2017**, *291*, 83-94. DOI: 10.1016/j.jneumeth.2017.07.031 From NLM Medline.

(18) Theis, L.; Berens, P.; Froudarakis, E.; Reimer, J.; Roman Roson, M.; Baden, T.; Euler, T.; Tolias, A. S.; Bethge, M. Benchmarking Spike Rate Inference in Population Calcium Imaging. *Neuron* **2016**, *90* (3), 471-482. DOI: 10.1016/j.neuron.2016.04.014 From NLM Medline.

(19) Sussillo, D.; Stavisky, S. D.; Kao, J. C.; Ryu, S. I.; Shenoy, K. V. Making brain-machine interfaces robust to future neural variability. *Nat Commun* **2016**, *7*, 13749. DOI: 10.1038/ncomms13749 From NLM Medline.

(20) Pandarinath, C.; O'Shea, D. J.; Collins, J.; Jozefowicz, R.; Stavisky, S. D.; Kao, J. C.; Trautmann, E. M.; Kaufman, M. T.; Ryu, S. I.; Hochberg, L. R.; et al. Inferring single-trial neural population dynamics using sequential auto-encoders. *Nat Methods* **2018**, *15* (10), 805-815. DOI: 10.1038/s41592-018-0109-9 From NLM Medline.

(21) Wen, X. Y.; Li, W. B. Time Series Prediction Based on LSTM-Attention-LSTM Model. *Ieee Access* **2023**, *11*, 48322-48331. DOI: 10.1109/Access.2023.3276628.

(22) Churchland, M. M.; Cunningham, J. P.; Kaufman, M. T.; Foster, J. D.; Nuyujukian, P.; Ryu, S. I.; Shenoy, K. V. Neural population dynamics during reaching. *Nature* **2012**, *487* (7405), 51-+. DOI: 10.1038/nature11129.

(23) Ioffe, S.; Szegedy, C. Batch Normalization: Accelerating Deep Network Training by Reducing Internal Covariate Shift. *Pr Mach Learn Res* **2015**, *37*, 448-456.

(24) Wu, Y. X.; He, K. M. Group Normalization. *Lect Notes Comput Sc* **2018**, *11217*, 3-19. DOI: 10.1007/978-3-030-01261-8_1.

(25) He, K. M.; Zhang, X. Y.; Ren, S. Q.; Sun, J. Deep Residual Learning for Image Recognition. *Proc Cvpr Ieee* **2016**, 770-778. DOI: 10.1109/Cvpr.2016.90.

(26) Churchland, M. M.; Shenoy, K. V. Preparatory activity and the expansive null-space. *Nat Rev Neurosci* **2024**, *25* (4), 213-236. DOI: 10.1038/s41583-024-00796-z From NLM Medline.

(27) Lin, T. Y.; Goyal, P.; Girshick, R.; He, K. M.; Dollár, P. Focal Loss for Dense Object Detection. *Ieee I Conf Comp Vis* **2017**, 2999-3007. DOI: 10.1109/Iccv.2017.324.

(28) Loshchilov, I.; Hutter, F. Decoupled Weight Decay Regularization. In *International Conference on Learning Representations*, 2017.

(29) Prechelt, L. Early Stopping — But When? In *Neural Networks: Tricks of the Trade: Second Edition*, Montavon, G., Orr, G. B., Müller, K.-R. Eds.; Springer Berlin Heidelberg, 2012; pp 53-67.

(30) Di Domenico, D.; Forsiuk, I.; Muller-Cleve, S.; Tanzarella, S.; Garro, F.; Marinelli, A.; Canepa, M.; Laffranchi, M.; Chiappalone, M.; Bartolozzi, C.; et al. Reach&Grasp: a multimodal dataset of the whole upper-limb during simple and complex movements. *Sci Data* **2025**, *12* (1), 233. DOI: 10.1038/s41597-025-04552-5 From NLM Medline.

(31) Dever, A.; Powell, D.; Graham, L.; Mason, R.; Das, J.; Marshall, S. J.; Vitorio, R.; Godfrey, A.; Stuart, S. Gait Impairment in Traumatic Brain Injury: A Systematic Review. *Sensors (Basel)* **2022**, *22* (4). DOI: 10.3390/s22041480 From NLM Medline.

(32) Park, S.; Lipton, M.; Dadarlat, M. C. Decoding multi-limb movements from two-photon calcium imaging of neuronal activity using deep learning. *J Neural Eng* **2024**, *21* (6). DOI: 10.1088/1741-2552/ad83c0 From NLM Medline.

(33) Zhang, X.; Landsness, E. C.; Miao, H.; Chen, W.; Tang, M.; Brier, L. M.; Culver, J. P.; Lee, J. M.; Anastasio, M. A. Attention-Based CNN-BiLSTM for Sleep State Classification of Spatiotemporal Wide-Field Calcium Imaging Data. *ArXiv* **2024**. From NLM PubMed-not-MEDLINE.